\documentstyle[prd,aps,preprint,epsfig]{revtex}
\begin{document}
%%%%%%%%%%%%%%%%%%%%%%%%%%%%%%%%%%%%%%%%%%%%%%%%%%%%%%%%
\newcommand{\TeV}{\,{\rm TeV}}
\newcommand{\GeV}{\,{\rm GeV}}
\newcommand{\MeV}{\,{\rm MeV}}
\newcommand{\keV}{\,{\rm keV}}
\newcommand{\eV}{\,{\rm eV}}
\def\ap{\approx}
\def\bea{\begin{eqnarray}}
\def\eea{\end{eqnarray}}
\def\bec{\begin{center}}
\def\ec{\end{center}}
\def\pC{\tilde{\chi}^+}
\def\nC{\tilde{\chi}^-}
\def\pnC{\tilde{\chi}^{\pm}}
\def\Ne{\tilde{\chi}^0}
\def\snu{\tilde{\nu}}
\def\tN{\tilde N}
\def\ler{\lesssim}
\def\gtr{\gtrsim}
\def\beq{\begin{equation}}
\def\eeq{\end{equation}}
\def\haf{\frac{1}{2}}
\def\plb#1#2#3#4{#1, Phys. Lett. {\bf #2B} (#4) #3}
\def\plbb#1#2#3#4{#1 Phys. Lett. {\bf #2B} (#4) #3}
\def\npb#1#2#3#4{#1, Nucl. Phys. {\bf B#2} (#4) #3}
\def\prd#1#2#3#4{#1, Phys. Rev. {\bf D#2} (#4) #3}
\def\prl#1#2#3#4{#1, Phys. Rev. Lett. {\bf #2} (#4) #3}
\def\mpl#1#2#3#4{#1, Mod. Phys. Lett. {\bf A#2} (#4) #3}
\def\rep#1#2#3#4{#1, Phys. Rep. {\bf #2} (#4) #3}
\def\lpp{\lambda''}
\def\ccg{\cal G}
\def\slash#1{#1\!\!\!\!\!/}
\def\rpv{\slash{R_p}}
\def\ler{\lesssim}
\def\gtr{\gtrsim}
\def\beq{\begin{equation}}
\def\eeq{\end{equation}}
\def\haf{\frac{1}{2}}
\def\plb#1#2#3#4{#1, Phys. Lett. {\bf #2B} (#4) #3}
\def\plbb#1#2#3#4{#1 Phys. Lett. {\bf #2B} (#4) #3}
\def\npb#1#2#3#4{#1, Nucl. Phys. {\bf B#2} (#4) #3}
\def\prd#1#2#3#4{#1, Phys. Rev. {\bf D#2} (#4) #3}
\def\prl#1#2#3#4{#1, Phys. Rev. Lett. {\bf #2} (#4) #3}
\def\mpl#1#2#3#4{#1, Mod. Phys. Lett. {\bf A#2} (#4) #3}
\def\rep#1#2#3#4{#1, Phys. Rep. {\bf #2} (#4) #3}
\def\lpp{\lambda''}
\def\ccg{\cal G}
\def\slash#1{#1\!\!\!\!\!/}
\def\rpv{\slash{R_p}}
\def\pslash{p\hspace{-2.0mm}/}
\def\qslash{q\hspace{-2.0mm}/}
\newcommand{\imag}{\Im {\rm m}}
\newcommand{\real}{\Re {\rm e}}

%%%%%%%%%%%%%%%%%%%%% Main Text %%%%%%%%%%%%%%%%
\setcounter{page}{1}
%\renewcommend{\arraystretch}{1.3}
\draft
%\widetext
\preprint{KAIST-TH 01/13, hep-ph/0108028}
\title{Sneutrino-Antisneutrino Mixing
and Neutrino Mass in Anomaly--mediated 
Supersymmetry Breaking Scenario}
\author{
Kiwoon Choi, Kyuwan Hwang, and Wan Young Song}
\address{Department of Physics,
Korea Advanced Institute of Science and Technology\\
Taejon 305-701, Korea}
\date{\today}
%
%\tighten
%
\maketitle
%
%%%%%%%%%%%%%%%%%%%%%%%%%%%%%%%%%%%%%%%%%%%%%%%%%%%%%%%%%%%%%
\begin{abstract}
%%%%%%%%%%%%%%%%%%%%%%%%%%%%%%%%%%%%%%%%%%%%%%%%%%%%%%%%%%%%%
In supersymmetric models with nonzero Majorana neutrino mass, the
sneutrino and antisneutrino mix, which may lead to same sign
dilepton signals in future collider experiments. We point out that
the anomaly-mediated supersymmetry breaking scenario has a good
potential to provide an observable rate of such signals for the
neutrino masses suggested by the atmospheric and solar neutrino
oscillations. The sneutrino mixing rate is naturally enhanced by
$m_{3/2}/m_{\tilde{\nu}}={\cal O}(4\pi/\alpha)$ while the
sneutrino decay rate is small enough on a sizable portion of the
parameter space. We point out also that the
sneutrino-antisneutrino mixing can provide much stronger
information on some combinations of the neutrino masses and mixing
angles than neutrino experiments.

%%%%%%%%%%%%%%%%%%%%%%%%%%%%%%%%%%%%%%%%%%%%%%%%%%%%%%%%%%%%%
\end{abstract}
%%%%%%%%%%%%%%%%%%%%%%%%%%%%%%%%%%%%%%%%%%%%%%%%%%%%%%%%%%%%%
%

Current data from the atmospheric and solar neutrino experiments
strongly suggest that the neutrinos have small but nonzero masses
\cite{numass}.
%Among the known,  the most natural way to generate
%a small neutrino mass is to assume that the neutrino mass
%is generated at high energy scale $M$ which is far above the weak
%scale, for instance by the exchange of heavy singlet neutrino
%as in the conventional seesaw model
%\cite{seesaw}.
%The resulting neutrino mass is then of Majorana-type and is small as
%it is suppressed by $1/M$.
As was pointed out in Refs.~\cite{Hirsch,GHaber,GHaber1,ejchun},
in supersymmetric models with  nonzero Majorana neutrino mass, the
sneutrino ($\snu$) and antisneutrino ($\snu^*$) mix to each other.
The mixing rate is generically given by \beq \Delta
m_{\tilde{\nu}}= C_\nu m_\nu/m_{\tilde{\nu}} \eeq where $m_\nu$ is
the lepton-number violating ($\Delta L=2$) Majorana neutrino mass,
$m_{\tilde{\nu}}$ is the lepton-number conserving ($\Delta L=0$)
sneutrino mass, and $C_\nu$ is determined mainly by the soft
supersymmetry (SUSY) breaking in the lepton-number violating
sector of the underlying theory.
%Depending on the size
%of $\Delta m_{\tilde{\nu}}$, we have in principle two possible
%ways to observe the $\snu$-$\snu^*$ mixing phenomena
%\cite{GHaber}. If $\Delta m_{\snu}$ is in the GeV range, the two
%mass-eigenstates $\snu_{\pm}=(\snu\pm\snu^*)/\sqrt{2}$ (for each
%sneutrino generation) would give distinguishable reasonant peaks
%in $e^+e^-$ annihilations. However
The atmospheric and solar neutrino data give the neutrino square
mass differences $\Delta m^2_{\nu}\lesssim {\cal O}(10^{-3}) \,
{\rm eV}^2$, so
%it is very unlikely
%that such a large $\Delta m_{\snu}$
%can be compatible with the atmospheric and solar neutrino oscillations
%\cite{numass}.
%The atmospheric neutrino data indicate a
%neutrino square mass difference
%$\Delta m^2_{\rm atm}\simeq 3\times 10^{-3} \, {\rm eV}^2$.
%Another square mass difference suggested by the solar neutrino
%data is even smaller, $\Delta m^2_{\rm sol}\,\lesssim \,{\cal O}(10^{-4}) \,
%{\rm eV}^2$.
it is quite unlikely that any of the neutrino masses is
significantly bigger than ${\cal O}(1)$ eV even when we include
the possibility of nearly degenerate neutrino masses. Also the
consideration of radiative corrections to $m_\nu$ induced by
$\Delta m_{\snu}$ \cite{GHaber} leads to the bound $\Delta
m_{\snu}/m_\nu\lesssim {\cal O}(4\pi/\alpha)$, implying $\Delta
{m}_{\snu} \, \lesssim \, {\cal O}(1)$ keV. Such a small mixing
rate can be probed by the $\snu$-$\snu^*$ oscillation which would
result in  same sign dilepton signals when the sneutrino pairs
decay into charged leptons.

To have an observable rate of same sign dilepton signals, $\snu$
must have an enough time to mix with $\snu^*$ before it decays.
For this, we need the sneutrino decay width $\Gamma_{\snu}\lesssim
{\cal O}(1)$ keV in view of $\Delta m_{\snu}\lesssim {\cal O}(1)$
keV. Such a small decay rate would not be possible if the two body
decay channel $\snu\rightarrow \nu\tilde{\chi}^0$ or
$\ell^-\tilde{\chi}^+$ is open for the neutralinos
$\tilde{\chi}^0$ or charginos $\tilde{\chi}^+$. It was pointed out
in \cite{GHaber} that the most plausible scenario for
$\Gamma_{\snu}\lesssim {\cal O}(1)$ keV is to have
\beq
\label{masshierarchy}
m_{\tilde{\tau}_1}\, < \, m_{\snu} \, < \, m_{\tilde{\chi}^0}, \,
m_{\tilde{\chi}^+}\,,
\eeq
where $\tilde{\tau}_1$ denotes the lighter stau.
Then sneutrinos decay mainly into  three-body final states
with sizable branching ratio into a charged lepton:
%\snu_{\ell} &\to& \pC_i \ell^-~(i=1,2),
%~~\snu_{\ell} \to \Ne_j \nu_{\ell}~(j=1-4)
$\snu \to \ell^- \tilde{\tau}_1^+ \nu_{\tau}, \nu
\tilde{\tau}_1^{\pm} \tau^{\mp} $ with $\Gamma_{\snu}\, \lesssim
\, {\cal O}(1)$ keV. The mass hierarchy (\ref{masshierarchy})
would mean that the stau is the lightest supersymmetric particle
in the minimal supersymmetric standard model (MSSM) sector, which
would not be cosmologically allowed if it is stable. This
difficulty can be easily avoided if one assumes a light singlet
fermion $\psi$  which has very weak couplings to the MSSM sector,
e.g. a light gravitino or axino, with which $\tilde{\tau}_1$
decays into $\tau\psi$. Alternatively, one may introduce a tiny
$R$-parity violating coupling
%$\lambda_{ijk} L_iL_j E^c_{k}$
which would trigger $\tilde{\tau}_1\rightarrow
\ell\nu$.
Still $\tilde{\tau}_1$ can live long  enough
inside the detector, so clearly distinguished from other
charged sleptons.
Then the same sign dilepton events induced by the
$\snu$-$\snu^*$ mixing are accompanied by long-lived same sign stau pairs,
so provide a rather clean signal for
the $\snu$-$\snu^*$ mixing if we focus on the final states
involving $\ell\ell\tilde{\tau}_1\tilde{\tau}_1$
for $\ell=e,\mu$.

Obviously, for a given neutrino mass, models with bigger
$C_\nu/m_{\tilde{\nu}}=\Delta m_{\snu}/m_\nu$ have better prospect
for observable $\tilde{\nu}$-$\tilde{\nu}^*$ mixing. In this
paper, we point out that the anomaly-mediated SUSY breaking (AMSB)
scenario \cite{amsb}  generically predicts
$C_\nu/m_{\tilde{\nu}}={\cal O}(4\pi/\alpha)\gg 1$ if the neutrino
mass is generated by SUSY preserving dynamics at high energy scale
as in the conventional seesaw model \cite{seesaw}. Furthermore the
mass hierarchy (\ref{masshierarchy}) can be obtained on a sizable
portion of the phenomenologically allowed parameter space of the
AMSB model. These features are not shared by the minimal
supergravity (SUGRA) model \cite{nilles} or the gauge-mediated
SUSY breaking (GMSB) model \cite{gmsb}, so the AMSB model has much
better potential to provide observable same sign dilepton signals
induced by the $\snu$-$\snu^*$ mixing than other SUSY breaking
models. An interesting feature of the $\snu$-$\snu^*$ mixing in
the AMSB scenario is that it provides rather strong information on
the neutrino mass matrix elements $(m_\nu)_{ii}=\sum_a
U^2_{ia}m_{\nu_a}$ where $i=e,\mu,\tau$ denote the flavor
eigenstates  while $a=1,2,3$ stand for the neutrino mass
eigenstates with the MNS mixing matrix $U_{ia}$. Atmospheric
neutrino data suggests $(m_\nu)_{\mu\mu} \simeq 3\times 10^{-2}$
eV  for  hierarchical neutrino masses, while $(m_\nu)_{\mu\mu}$
can be bigger if the neutrino masses are approximately degenerate.
Then the same sign dimuon events from
 $\snu_\mu$-$\snu^*_\mu$ mixing can be used to distinguish
$(m_\nu)_{\mu\mu}=3\times 10^{-2}$ eV from a bigger value of
$(m_\nu)_{\mu\mu}$. Also the same sign dielectron events from
$\snu_e$-$\snu^*_e$ mixing can be used to probe $(m_\nu)_{ee}$
down to the order of $10^{-4}$ eV which is much smaller than the
current bound on $(m_\nu)_{ee}$ from $\nu$-less double beta
decays. In the following, we will examine these points in more
detail.

Let us first discuss  some generic features of the AMSB model.
Anomaly mediation assumes that supersymmetry breaking in the
hidden sector is transmitted to the MSSM fields {\it mainly} by
the Weyl compensator superfield $\Phi_0$ of the supergravity
multiplet \cite{amsb}: \beq \label{compensator} \Phi_0=1+\theta^2
M_{\rm aux}, \eeq where $M_{\rm aux}$ is generically of order the
gravitino mass $m_{3/2}$. The couplings of $\Phi_0$ to generic
matter multiplets are determined  by the super-Weyl invariance.
Therefore at classical level, $\Phi_0$ is coupled to the MSSM
fields only through {\it dimensionful} supersymmetric parameters,
while the $\Phi_0$-couplings through dimensionless parameters
arise from radiative corrections. More explicitly, a super-Weyl
invariant effective action can be written as \bea S_{\rm
eff}&=&\int \, d^4x d^4\theta \, \left[\, {Z}_{_I}
(Q/\sqrt{\Phi_0\Phi_0^*}) \Phi_{_I}^*\Phi_{_I}
+\frac{1}{8}g^{-2}_a(Q/\sqrt{\Phi_0\Phi_0^*})
DV^a\bar{D}^2DV^a\,\right]
\nonumber \\
&+&\left[\,\int \, d^4xd^2\theta \, \left(\,
y_{_{IJK}}\Phi_{_I}\Phi_{_J} \Phi_{_K}
+\frac{\Phi_0}{M}\gamma_{_{IJKL}}\Phi_{_I}
\Phi_{_J}\Phi_{_K}\Phi_{_L}\, \right) +{\rm h.c.}\,\right] \eea
where $Q$ denotes the renormalization scale, $D$ and $\bar{D}$ are
the supercovariant derivatives on the real gauge superfields
$V_a$, and $\Phi_{_I}$ are the chiral matter superfields. Here the
quartic terms  in the superpotential are assumed to be induced by
supersymmetry preserving dynamics, e.g. by the exchange of heavy
particles with supersymmetric mass $M$. Note that such heavy
particles can be integrated out {\it while preserving the
super-Weyl invariance}.

With $\Phi_0$ given as (\ref{compensator}),
$S_{\rm eff}$ would
determine the pure anomaly-mediated soft parameters
in a manner which is valid at an arbitrary scale $Q$.
However it predicts a tachyonic slepton,
so one needs some additional source of SUSY breaking.
The simplest possibility is
an additional  universal soft scalar square mass
$m_0^2$ introduced  at the GUT scale $M_{\rm GUT}\simeq 2\times 10^{16}$ GeV.
This defines the minimal AMSB model which predicts
the following forms of soft supersymmetry
breaking terms \cite{amsb,amsb1}:
\beq
{\cal L}_{\rm soft}=m^2_{_I}|\phi_{_I}|^2
+\left(\frac{1}{2}M_a\lambda^a\lambda^a+
A_{_{IJK}}y_{_{IJK}}\phi_{_I}\phi_{_J}\phi_{_K}
+\frac{C_{_{IJKL}}\gamma_{_{IJKL}}}{M}\phi_{_I}\phi_{_J}
\phi_{_K}\phi_{_L}
+{\rm h.c.}\right)
\eeq
where the gaugino masses $M_a$,
the soft scalar masses $m_{_I}$ and
the soft coefficients $A_{_{IJK}}, C_{_{IJKL}}$ are given by
\bea
\label{amsbbc}
&& M_a
=-\frac{b_a\alpha_a}{4\pi} M_{\rm aux},
\quad m^2_{_I}=
-\frac{1}{4}\frac{d\gamma_{_I}}{d\ln Q}|M_{\rm aux}|^2+m_0^2,
\nonumber \\
&& A_{_{IJK}}
=\frac{1}{2}(\gamma_{_I}+\gamma_{_J}+\gamma_{_K})M_{\rm aux},
\nonumber \\
&& C_{_{IJKL}}=\frac{1}{2}(2+\gamma_{_I}+\gamma_{_J}+
\gamma_{_K}+\gamma_{_L})M_{\rm aux}. \eea Here $b_a=(3,-1,-33/5)$
are the one-loop beta function coefficients for $SU(3)_c\times
SU(2)_L\times U(1)_Y$ in the GUT normalization and $\gamma_{_I}=d
\ln Z_{_I}/d\ln Q$ are the anomalous dimension of $\Phi_{_I}$.
Note that still the expressions of $M_a$, $A_{_{IJK}}$ and
$C_{_{IJKL}}$ are valid at arbitrary scale, while the expression
of $m_{_I}^2$ is valid {\it only} at $M_{\rm GUT}$.

Applying the above results to the
sneutrino-antisneutrino mixing is rather straightforward.
To be specific, we will assume that the neutrino masses are
generated (mainly) by supersymmetry
preserving dynamics at an energy scale $M$ far above the
weak scale. This high energy dynamics may be the
exchange of heavy singlet neutrino with mass $M$
\cite{seesaw}, or the exchange of heavy
triplet Higgs boson \cite{triplet}, or some stringy dynamics.
Independently of its detailed shape,
this high energy dynamics can be integrated out while preserving
the super-Weyl invariance.
Then at the weak scale,
the theory can be described  by an effective superpotential
including the super-Weyl invariant dimension 5 operators for neutrino masses
and also the associated soft SUSY breaking terms,
\bea
\label{superpotential}
&& \Delta W_{\rm eff}=
\frac{\Phi_0}{M}\gamma_{ij}(L_iH_2)(L_jH_2),
\nonumber \\
&& \Delta {\cal L}_{\rm soft} =
\frac{C_{ij}\gamma_{ij}}{M}(\tilde{\ell}_ih_2)(\tilde{\ell}_jh_2),
\eea
where $L_i$ ($i=e,\mu,\tau$) and $H_\alpha$ ($\alpha=1,2$) denote
the lepton and Higgs  doublet superfields with the scalar
components $\tilde{\ell}_i$ and $h_\alpha$, respectively, and
$C_{ij}\simeq M_{\rm aux}$.

After the electroweak symmetry breaking,
$\Delta W_{\rm eff}$ gives a neutrino mass matrix
$(m_\nu)_{ij}=2 \langle h_2\rangle^2\gamma_{ij}/M$.
%This $m^\nu$ can be diagonalized as
%$U^Tm^\nu\, U=
%{\rm diag}\, (m_{\nu_1}, m_{\nu_2}, m_{\nu_3})$
%where $U_{ia}$ ($a=1,2,3$) corresponds to the MNS lepton flavor mixing matrix
%when $L_i$ are chosen to be the charged lepton mass eigenstates.
Including the contribution from $\Delta {\cal L}_{\rm soft}$, the
sneutrino masses are given by \bea
(m^2_{\tilde{\nu}})_{ij}\snu_i^*\tilde{\nu}_j
+\left\{\frac{1}{2}(\Delta m^2_{\tilde{\nu}})_{ij}\tilde{\nu}_i
\tilde{\nu}_j+{\rm h.c.} \right\}\, , \eea where the sneutrino
square mass matrix $(m^2_{\tilde{\nu}})_{ij}\simeq
m_{\snu}^2\delta_{ij}$ with $m_{\snu}^2=\frac{1}{2}M_Z^2\cos
2\beta + m^2_{\tilde{\ell}}$ for the slepton doublet square mass
matrix $(m_{\tilde{\ell}}^2)_{ij}\simeq
m^2_{\tilde{\ell}}\delta_{ij}$ and the $\Delta L=2$ sneutrino
square mass matrix is given by $(\Delta m^2_{\tilde{\nu}})_{ij}
=(C_{ij} +2\mu \cot\beta)(m_\nu)_{ij}$.
%In the AMSB model,
%both $(m_{\tilde{\nu}}^2)_{ij}$ and $C_{ij}$ at the weak scale
%are flavor-independent up to small Yukawa-induced
%corrections:
%\beq
%(m^2_{\tilde{\nu}})_{ij}\simeq m^2_{\tilde{\nu}} \delta_{ij}
%={\cal O}(\frac{\alpha}{4\pi}M_{\rm aux})
%\quad
%C_{ij}\simeq C_\nu\delta_{ij}.
%\eeq
The $\tilde{\nu}$-$\tilde{\nu}^*$ mixing rate is determined by the
sneutrino mass-splitting $\Delta m_{\snu}=\Delta
m^2_{\snu}/m_{\snu}$. In this regard, a {\it distinctive feature}
of the AMSB model is that $C_{ij}\simeq M_{\rm aux}$ is induced at
tree level while $m_{\snu}$ is loop-suppressed, so $\Delta
m_{\snu}$ is enhanced (relative to $m_\nu$) by the factor $M_{\rm
aux} /m_{\snu} ={\cal O}(4\pi/\alpha)$: \beq
\label{snumasssplitting} (\Delta m_{\tilde{\nu}})_{ij}\simeq
 (m_\nu)_{ij}M_{\rm aux}/m_{\snu}={\cal O}(4\pi m_\nu/\alpha).
\eeq
Consequently, for a given neutrino mass,
 the AMSB model has better potential
to give a sizable $\snu$-$\snu^*$  mixing than other models with
$C_{ij}/m_{\snu}={\cal O}(1)$. Furthermore, as can be inferred
from Fig. 1, a significant portion of the phenomenologically
viable parameter space of the minimal AMSB model leads to the mass
hierarchy (\ref{masshierarchy}), which is a feature {\it not}
shared by the minimal SUGRA or GMSB models. This is partly because
in the AMSB model the lightest neutralino ($\tilde{\chi}_1^0$) and
the lightest chargino ($\pnC_1$) are nearly degenerate and the
mass gap between the sleptons and
$\tilde{\chi}_1^0,\tilde{\chi}_1^+$ is  narrower than the minimal
SUGRA and GMSB models \cite{amsb1}.

Taking an analogy to the $B$-meson mixing, it is straightforward
to compute the probability for a $\snu$-$\snu^*$ pair produced in
$e^+e^-$ collider to yield same-sign dilepton signal
\cite{GHaber,chs}. The amplitude for $e^+(p_1) + e^-(p_2) \to
  \tilde{\nu}_i(q_1) + \tilde{\nu}_i^*(q_2)$ is easily computed to
be
\beq
A_i=\frac{1}{2} g^2 \bar{v}(p_2) (\qslash_1 - \qslash_2)
             (X_iP_L + Y_iP_R) u(p_1)\,,
\eeq
where $P_{L,R}=(1\pm\gamma_5)/2$,
$X_i=K_Z(s_W^2-\frac{1}{2})/c_W^2
      + \delta_{ie} \sum_n|V_{n1}|^2K_n$,
$Y_i=K_Z s_W^2/c_W^2$ for the chargino ($\tilde{\chi}^{\pm}_n$)
mixing matrix $V_{nm}$, $K_Z =1/(s-M_Z^2)$, $K_n = 1/[(p_1-q_1)^2
- m^2_{\chi_n}]$, $c_W=\cos\theta_W$, $s_W=\sin\theta_W$, and the
c.m. energy $\sqrt{s}$. Here $K_Z$ represents the contributions
from the $Z$ boson mediated $s$-channel diagrams, while $K_n$ is
from the chargino mediated $t$-channel diagrams.
%Among the  produced sneutrino states,
%two states with the momentum pointing the opposite direction of the other one
% can interfere with each other by sneutrino-antisneutrino mixing.
With this amplitude, the initial $\snu$-$\snu^*$ state is
given by
\beq
\label{initial}
|\snu \snu^*;0 \rangle
= \sum_i \, \alpha_i |\snu_{i}(\vec{q})\rangle
|\snu_{i}^*(-\vec{q})\rangle
    +  \beta_i |\snu_{i}^{*}(\vec{q})\rangle
|\snu_i(-\vec{q})\rangle\, , \eeq where
$\alpha_i=A_i(\vec{q_1}=\vec{q})$,
$\beta_i=A_i(\vec{q_2}=\vec{q})$ (up to overall normalization) and
the momentum vector $\vec{q}$ spans only the upper hemisphere,
i.e. $\cos \theta \geq 0$ for the angle $\theta$ between $e^-$ and
$\tilde{\nu}$ flight directions.
%$(p_1-q_1)^2 = m^2_{\tilde{\nu}}-s(1-\beta\cos\theta)$
%where $\beta = (1-4m^2_{\tilde{\nu}}/s)^{1/2}$.
%\beq
%\sum_i |\alpha_i|^2 + |\beta_i|^2 = 1\,.
%\eeq

With (\ref{masshierarchy}), the sneutrinos decay as
$\snu\to\ell^-\tilde{\tau}_1^+\nu_{\tau},
\nu\tilde{\tau}_1^{\pm}\tau^{\mp}$. It turns out that, in most of
the parameter space yielding an observable rate of same sign
dilepton signals,
%$\tan\beta$ is not large and
%$m_{\snu_{\ell}}-m_{\tilde{\tau}_1}\gg m_{\tau}$, so
these decays are induced dominantly by the chargino or neutralino
exchange, so the decay widths are (approximately)
flavor-independent. Then the effective Hamiltonian determining the
evolution (\ref{initial}) can be written as
 \beq  \label{effH} H_{\rm eff} = \pmatrix{
(m_{\snu}-{i\over 2}\Gamma_{\snu})\delta_{ij}
       + (\delta m_{\snu})_{ij}
       & {1\over 2} (\Delta m_{\snu})_{ij} \cr
       {1\over 2} (\Delta m_{\snu})_{ij}^* &
         (m_{\snu}-{i\over 2}\Gamma_{\snu})\delta_{ij}
       + (\delta m_{\snu})_{ij}^*   }\,,
\eeq where $\delta m_{\snu}$ represents the deviation from the
exact degeneracy of the $\Delta L=0$ sneutrino masses. Since
$\delta m_{\snu}\gg \Delta m_{\snu}$ in our case, it is most
convenient  to describe the $\snu$-$\snu^*$ mixing in the field
basis in which $\delta m_{\snu}$ is diagonal. In the AMSB
scenario, the charged lepton mass matrix can be diagonalized
simultaneously with $\delta m_{\snu}$. In such field basis, we
find the probability $P_i$ for the initial state (\ref{initial})
to produce a same sign dilepton $\ell_i^- \ell_i^-$ or
 $\ell_i^+\ell_i^+$:
\bea \label{probability} P_i  &=& \frac{1}{\sum_i\sigma_i} \int
d\Phi_2\frac{1}{8s}
     {B_i^2 \over (1+x_i^2)^2} \left\{
     \frac{1}{2}(|\alpha_i|^2+|\beta_i|^2)(2+x_i^2)x_i^2
     + {\rm Re}(\alpha^*_i \beta_i) x_i^2 \right\},
\eea where $B_i={\rm Br}(\snu_i\rightarrow \ell_i X)$, $x_i
=|(\Delta m_{\snu})_{ii}|/\Gamma_{\snu} = |(m_\nu)_{ii}|M_{\rm aux}/
m_{\snu}\Gamma_{\snu}$ and  $\sigma_i$ denotes the total cross
section for $e^+e^-\rightarrow\snu_i\snu_i^*$. Here the 2-body
phase space integration ($d\Phi_2$) for the initial $\snu\snu^*$
is performed for $\cos \theta\ge 0$ and we have ignored the
effects suppressed by $\Delta m_{\snu}/\delta m_{\snu}$.

The same sign dilepton probability (\ref{probability}) shows that
in the AMSB scenario the $\snu$-$\snu^*$ mixing provides
information on the neutrino matrix elements
$(m_\nu)_{ii}=\sum_aU^2_{ia} m_{\nu_a}$ where $m_{\nu_a}$ and
$U_{ia}$ denote the neutrino mass eigenvalues and the MNS mixing
matrix, respectively. This would be true in other SUSY breaking
models as long as the SUSY breaking is transmitted to the
observable sector in a {\it flavor-blind} way. Currently
$(m_\nu)_{ee}$ is bounded to be less than 0.2 eV  by the
$\nu$-less double beta decay and this bound can be relaxed by  a
factor of few due to the uncertainty in the involved nuclear
matrix elements. As we will see, the $\snu$-$\snu^*$ mixing allows
us to probe $(m_\nu)_{ee}$ down to the order of $10^{-4}$ eV in
the AMSB scenario. Various neutrino oscillation experiments
including the atmospheric and solar neutrino oscillations provide
information on $m^2_{\nu_a}-m^2_{\nu_b}$ and $U_{ia}$,  for
instance $\Delta m^2_{\rm atm}=|m^2_{\nu_3}-m^2_{\nu_2}|^2\simeq 3
\times 10^{-3} \, {\rm eV}^2$ and $|U_{\mu 3}|\simeq |U_{\tau 3}|
=1/\sqrt{2}$. Still the information on $(m_\nu)_{ii}$ from the
$\snu$-$\snu^*$ mixing are different from the information on
neutrino masses and mixing angles  from neutrino oscillations. So
the $\snu$-$\snu^*$ mixing can provide information on neutrino
masses and mixing angles which are complementary to  those from
the neutrino experiments.

Same sign dilepton events may come also from the pair-produced
neutralinos which would decay as $\tilde{\chi}^0\rightarrow
\ell\tilde{\ell}$. However as long as we focus on the $e$ and
$\mu$ flavors the same sign dileptons from the neutralino pair
accompany $\tilde{e}$ or $\tilde{\mu}$, so can be clearly
distinguished from those from the $\snu$-$\snu^*$ mixing
accompanying the $\tilde{\tau}_1$ pair. With this observation, we
performed a numerical analysis to find the parameter region of the
minimal AMSB model yielding a sizable number of same sign dilepton
events per year,  $N_i$ ($i=e,\mu$), for a future $e^+e^-$ linear
collider with the integrated luminosity 500 ${\rm fb}^{-1}$ at
$\sqrt{s}=500$ GeV. As usual, we replace the Higgs $\mu$ and $B$
parameters by $M_Z$ and $\tan\beta$ under the condition of
electroweak symmetry breaking. Using the standard RG analysis, the
superparticle mass spectrums are obtained to compute
$\Gamma_{\snu}$ on the parameter region of $(m_0,M_{\rm
aux},\tan\beta)$ leading to the mass hierarchy
(\ref{masshierarchy}). When $\tan\beta$ increases for a given
$M_{\rm aux}$, the mass hierarchy (\ref{masshierarchy}) requires a
larger $m_0$ leading to a larger $m_{\snu}$. On the other hand,
the enhanced left-right mixing gives an effect to lower
$m_{\tilde{\tau}_1}$, so the net result is to increase
$m_{\snu}/m_{\tilde{\tau}_1}$. The  phase space of the three body
decays \, $\snu\to \ell \tilde{\tau}_1\nu_{\tau},\nu
\tilde{\tau}_1\tau$ \, is highly sensitive to
$m_{\snu}/m_{\tilde{\tau}_1}$. As a result, for a given neutrino
mass, $\Gamma_{\snu}$ is a  sharply increasing function of
$\tan\beta$, so small $\tan\beta$ is favored for sizable $N_i$.
From a detailed numerical analysis, we find that
$\tan\beta\lesssim 10$ is required to have a sizable $N_i$ for
$m_\nu\lesssim {\cal O}(1)$ eV.

About the values of $(m_\nu)_{\mu\mu}$, we considered two cases.
In the first,  neutrino masses  are assumed to be hierarchical,
which would give $(m_\nu)_{\mu\mu}\simeq U^2_{\mu 3}\sqrt{\Delta
m^2_{\rm atm}}\simeq 3\times 10^{-2}$ eV, while in the second case
neutrino masses  are assumed to be approximately degenerate with
$(m_\nu)_{\mu\mu}=0.3$ eV.  We find that the number of same sign
dimuon events per year ($N_\mu$) for the first case is bigger than
the value for the second case by about factor 5, so hierarchical
and (approximately) degenerate neutrino masses are clearly
distinguished from each other. In Fig. 1, we depict the parameter
regions  with  $\tan\beta=5$ yielding $N_\mu\geq 20, 10^2, 5\times
10^2$ for $(m_\nu)_{\mu\mu}= 3\times 10^{-2}$ eV  and $N_\mu\geq
10^2, 5\times 10^2, 2\times 10^3$ for $(m_\nu)_{\mu\mu}=0.3$ eV.
 We also searched for the parameter regions with $\tan\beta=5$ yielding
 $N_{e}\geq 20, 10^2, 5\times 10^2$
 for $(m_\nu)_{ee}=10^{-2}, 10^{-3},10^{-4}$ eV and depict the results in Fig. 2.
 Note that the $t$-channel contribution to
$e^+e^-\rightarrow\snu_e\snu^*_e$ enhances $N_e$ relative to
$N_\mu$, so that we can have a sizable $N_e$ even for $(m_{ee})
=10^{-4}$ eV.

 As available constraints on
the model, we impose the Higgs, stau, chargino mass bounds:
$m_h>113.5$ GeV, $m_{\tilde{\tau}}>89$ GeV,
 $m_{\pnC_1}>103$ GeV, and also the $2\sigma$ constraint
on the $b \to s \gamma$ branching ratio:
${\rm Br}(B \to X_s \gamma)=(2.2- 4.1) \times 10 ^{-4}$.
It has been noted that the AMSB model is severely constrained
by the recent measurement of the muon anomalous magnetic moment
$a_\mu$  once we require that
the conventional one-loop SUSY contribution
$a_\mu^{\rm SUSY}\gtrsim 10^{-9}$ which was taken
as the $2\sigma$ lower bound \cite{feng}.
Here we do {\it not} take this as a real constraint since
the uncertainty in the hadronic contributions to $a_\mu$
can be as large as $10^{-9}$ \cite{yndu}.
Although not taken as a constraint,
we specify the parameter region
with $a_\mu^{\rm SUSY}\gtrsim 10^{-9}$ or $5\times 10^{-10}$
for the completeness.

To conclude, we have examined the possibility of an observable
same sign dilepton signal induced by the $\snu$-$\snu^*$ mixing in
the AMSB model. It is pointed out that the AMSB model has a good
potential to provide an observable rate of signals since the
mixing rate is naturally enhanced by $m_{3/2}/m_{\snu}={\cal
O}(4\pi/\alpha)$ while the sneutrino decay rate is small enough on
the sizable portion of the phenomenologically allowed parameter
space of the model. Our results depicted in Figs. 1 and 2 show
that this is indeed the case for the neutrino masses suggested by
the atmospheric and solar neutrino data. It is noted also that the
same-sign dilepton signals can be used to determine $(m_\nu)_{ee}$
and $(m_\nu)_{\mu\mu}$, providing useful information on the
neutrino masses and mixing angles which are complementary to those
from neutrino experiments.
% the resulting $P_{e\mu}$ would
%be suppressed by either $|U_{e2}|^2\simeq 10^{-3}$ or $|U_{e3}|^2
%\lesssim 4\times 10^{-2}$, so can be distinguished from
%the LMA result.

%
\bigskip

%%%%%%%%%%%%%%%%%%%%%%%%%%%%%%%%%%%%%%%%%%%%%%%%%%%%%%%%%
{\bf Acknowledgments}:
%%%%%%%%%%%%%%%%%%%%%%%%%%%%%%%%%%%%%%%%%%%%%%%%%%%%%%%%%
We thank E. J. Chun and P. Ko for useful discussions.
This work is supported by the BK21 project of the Ministry
of Education, KRF Grant No. 2000-015-DP0080, KOSEF Grant
No. 2000-1-11100-001-1, and the Center for High Energy Physics
of Kyungbook National University.
%%%%%%%%%%%%%%%%%%%%%%%%%%%%%%%%%%%%%%%%%%%%%%%%%%%%%%%%%%%

%%%%%%%%%%%%%%%%%%%%%%%%%%%%%%%%%%%%%%%%%%%%%%%%%%%%%%%%%%%
%
\begin{figure}
\begin{center}
\epsfig{file=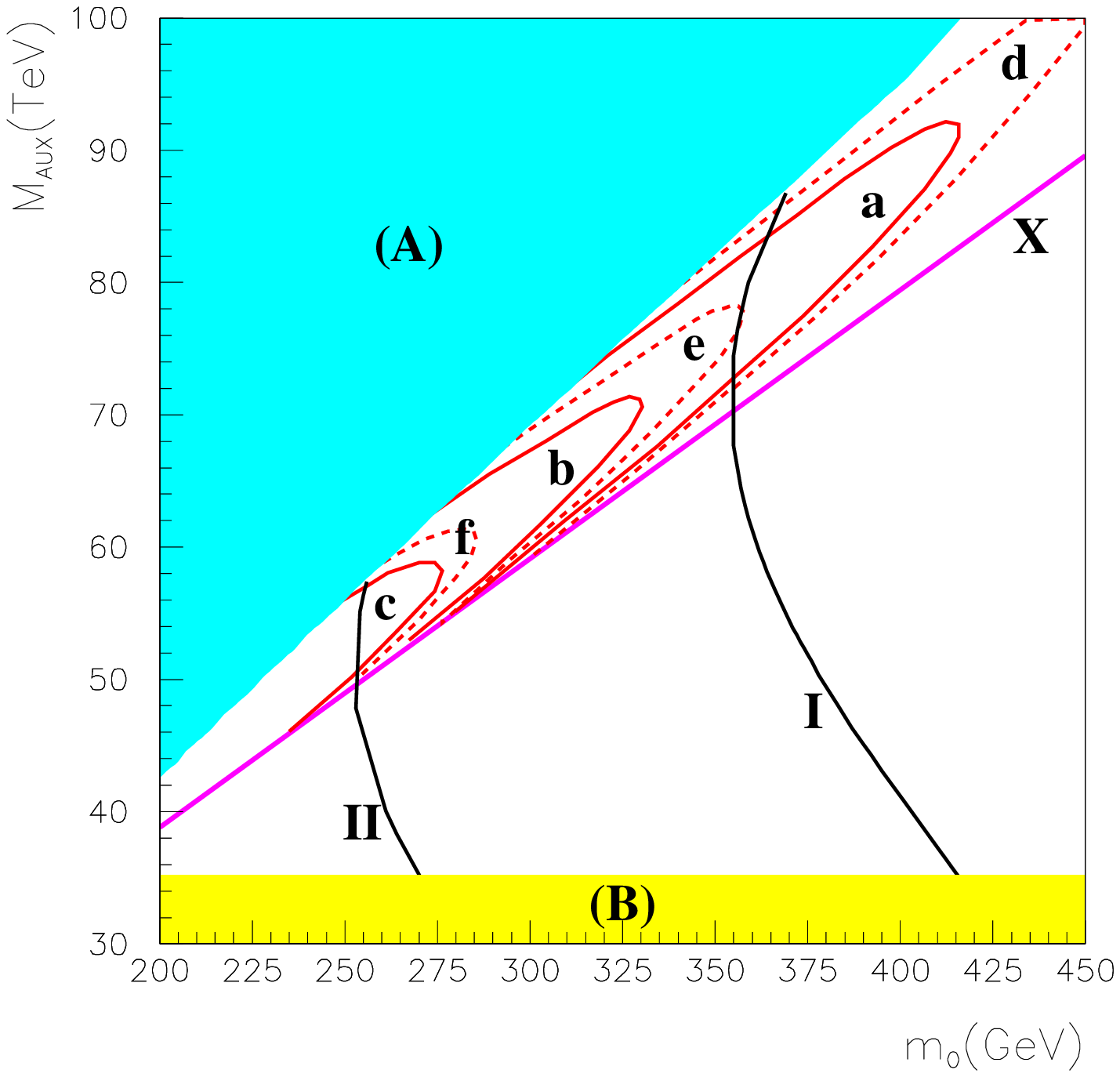,width=17cm,height=16cm}
\end{center}
\caption{Parameter regions with $\tan\beta=5$ yielding
$N_{\mu}\geq 20$
 [inside the contour $a$], $10^2$ $[b]$, $5\times 10^2$
$[c]$ for $(m_\nu)_{\mu\mu}=3\times 10^{-2}$  eV; $N_{\mu}\geq
10^2$ $[d]$, $5\times 10^2$ $[e]$, $2\times 10^3$ $[f]$ for
$(m_\nu)_{\mu\mu}=0.3$ eV.  (A) and (B) represent the parameter
regions forbidden by the stau and chargino mass bounds,
respectively. Upper side of the line $X$ denotes the region of LSP
stau. Left sides of the lines (I) and (II) correspond to the
region with $a_{\mu}^{\rm SUSY} \geq 5\times 10^{-10}$ and
$10^{-9}$, respectively.}
%\label{fig:possible}
%
\begin{center}
\epsfig{file=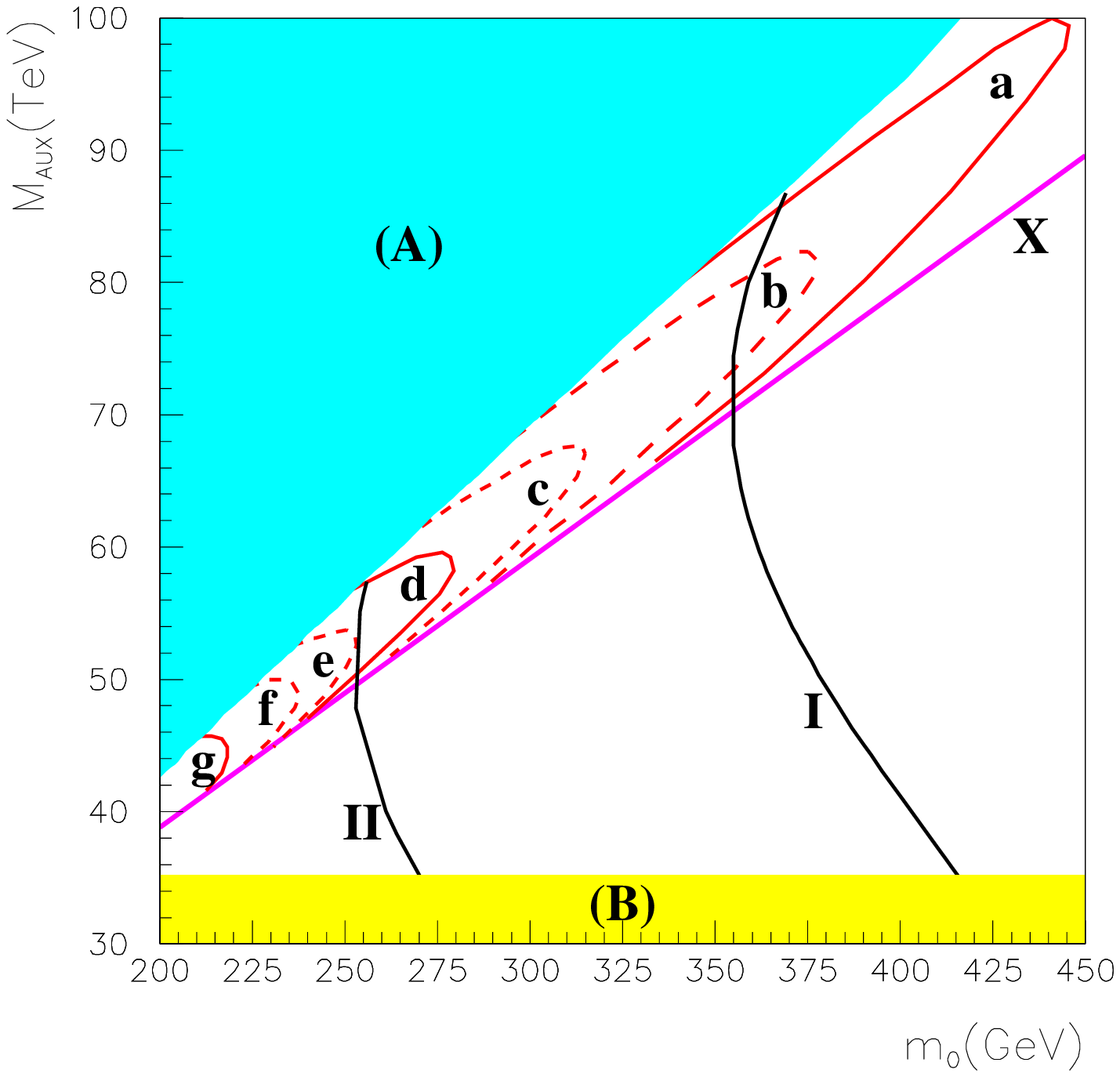,width=17cm,height=16cm}
\end{center}
\caption{Parameter regions with $\tan\beta=5$ yielding $N_e\geq
20$ $[a]$, $10^2$ $[b]$, $5\times 10^2$ $[c]$ for
$(m_\nu)_{ee}=10^{-2}$ eV; $N_e\geq 20$ $[d]$, $10^2$ $[e]$,
$5\times 10^2$ $[f]$ for $(m_\nu)_{ee}=10^{-3}$ eV; $N_e\geq 10^2$
$[g]$ for $(m_\nu)_{ee}=10^{-4}$ eV. }
%\label{fig:possible}
%
\end{figure}
%
%%%%%%%%%%%%%%%%%%%%%%%%%%%%%%%%%
\end{document}